
\def\singlespace{\normalbaselines}
\def\oneandahalfspace{\baselineskip=1.15\normalbaselineskip plus 1pt
\lineskip=2pt\lineskiplimit=1pt}

\def\np{\vfill\eject}
\def\nl{\hfil\break}

\def\nofirstpagenoten{\nopagenumbers\footline={\ifnum\pageno>1\tenrm
\hss\folio\hss\fi}}
\def\nofirstpagenotwelve{\nopagenumbers\footline={\ifnum\pageno>1\twelverm
\hss\folio\hss\fi}}
\def\leaderfill{\leaders\hbox to 1em{\hss.\hss}\hfill}
\def\ft#1#2{{\textstyle{{#1}\over{#2}}}}
\def\frac#1/#2{\leavevmode\kern.1em
\raise.5ex\hbox{\the\scriptfont0 #1}\kern-.1em/\kern-.15em
\lower.25ex\hbox{\the\scriptfont0 #2}}
\def\sfrac#1/#2{\leavevmode\kern.1em
\raise.5ex\hbox{\the\scriptscriptfont0 #1}\kern-.1em/\kern-.15em
\lower.25ex\hbox{\the\scriptscriptfont0 #2}}


\parindent=20pt
\def\narrow{\advance\leftskip by 40pt \advance\rightskip by 40pt}

\def\AB{\bigskip
        \centerline{\bf ABSTRACT}\medskip\narrow}
\def\nonarrower{\advance\leftskip by -40pt\advance\rightskip by -40pt}
\def\AE{\bigskip\nonarrower}

\def\boxit#1{\vbox{\hrule\hbox{\vrule\kern3pt
        \vbox{\kern3pt#1\kern3pt}\kern3pt\vrule}\hrule}}

\def\gtorder{\mathrel{\raise.3ex\hbox{$>$}\mkern-14mu
             \lower0.6ex\hbox{$\sim$}}}
\def\ltorder{\mathrel{\raise.3ex\hbox{$<$}|mkern-14mu
             \lower0.6ex\hbox{\sim$}}}
\def\dalemb#1#2{{\vbox{\hrule height .#2pt
        \hbox{\vrule width.#2pt height#1pt \kern#1pt
                \vrule width.#2pt}
        \hrule height.#2pt}}}

\font\fourteentt=cmtt10 scaled \magstep2
\font\fourteenbf=cmbx12 scaled \magstep1
\font\fourteenrm=cmr12 scaled \magstep1
\font\fourteeni=cmmi12 scaled \magstep1
\font\fourteenss=cmss12 scaled \magstep1
\font\fourteensy=cmsy10 scaled \magstep2
\font\fourteensl=cmsl12 scaled \magstep1
\font\fourteenex=cmex10 scaled \magstep2
\font\fourteenit=cmti12 scaled \magstep1
\font\twelvett=cmtt10 scaled \magstep1 \font\twelvebf=cmbx12
\font\twelverm=cmr12 \font\twelvei=cmmi12
\font\twelvess=cmss12 \font\twelvesy=cmsy10 scaled \magstep1
\font\twelvesl=cmsl12 \font\twelveex=cmex10 scaled \magstep1
\font\twelveit=cmti12
\font\tenss=cmss10
 
 \font\ninebf=cmbx7 scaled \magstep1
\font\ninerm=cmr7 scaled \magstep1 \font\ninei=cmmi7 scaled \magstep1
\font\ninesy=cmsy7 scaled \magstep1 
\font\eightrm=cmr7 scaled 1140 
 
\font\sevenbf=cmbx7 \font\sevenrm=cmr7 \font\seveni=cmmi7
\font\sevensy=cmsy7 

\catcode`@=11
\newskip\ttglue
\newfam\ssfam

\def\fourteenpoint{\def\rm{\fam0\fourteenrm}
\textfont0=\fourteenrm \scriptfont0=\tenrm \scriptscriptfont0=\sevenrm
\textfont1=\fourteeni \scriptfont1=\teni \scriptscriptfont1=\seveni
\textfont2=\fourteensy \scriptfont2=\tensy \scriptscriptfont2=\sevensy
\textfont3=\fourteenex \scriptfont3=\fourteenex \scriptscriptfont3=\fourteenex
\def\it{\fam\itfam\fourteenit} \textfont\itfam=\fourteenit
\def\sl{\fam\slfam\fourteensl} \textfont\slfam=\fourteensl
\def\bf{\fam\bffam\fourteenbf} \textfont\bffam=\fourteenbf
\scriptfont\bffam=\tenbf \scriptscriptfont\bffam=\sevenbf
\def\tt{\fam\ttfam\fourteentt} \textfont\ttfam=\fourteentt
\def\ss{\fam\ssfam\fourteenss} \textfont\ssfam=\fourteenss
\tt \ttglue=.5em plus .25em minus .15em
\normalbaselineskip=16pt
\abovedisplayskip=16pt plus 4pt minus 12pt
\belowdisplayskip=16pt plus 4pt minus 12pt
\abovedisplayshortskip=0pt plus 4pt
\belowdisplayshortskip=9pt plus 4pt minus 6pt
\parskip=5pt plus 1.5pt
\setbox\strutbox=\hbox{\vrule height12pt depth5pt width0pt}
\let\sc=\tenrm
\let\big=\fourteenbig \normalbaselines\rm}
\def\fourteenbig#1{{\hbox{$\left#1\vbox to12pt{}\right.\n@space$}}}

\def\twelvepoint{\def\rm{\fam0\twelverm}
\textfont0=\twelverm \scriptfont0=\ninerm \scriptscriptfont0=\sevenrm
\textfont1=\twelvei \scriptfont1=\ninei \scriptscriptfont1=\seveni
\textfont2=\twelvesy \scriptfont2=\ninesy \scriptscriptfont2=\sevensy
\textfont3=\twelveex \scriptfont3=\twelveex \scriptscriptfont3=\twelveex
\def\it{\fam\itfam\twelveit} \textfont\itfam=\twelveit
\def\sl{\fam\slfam\twelvesl} \textfont\slfam=\twelvesl
\def\bf{\fam\bffam\twelvebf} \textfont\bffam=\twelvebf
\scriptfont\bffam=\ninebf \scriptscriptfont\bffam=\sevenbf
\def\tt{\fam\ttfam\twelvett} \textfont\ttfam=\twelvett
\def\ss{\fam\ssfam\twelvess} \textfont\ssfam=\twelvess
\tt \ttglue=.5em plus .25em minus .15em
\normalbaselineskip=14pt
\abovedisplayskip=14pt plus 3pt minus 10pt
\belowdisplayskip=14pt plus 3pt minus 10pt
\abovedisplayshortskip=0pt plus 3pt
\belowdisplayshortskip=8pt plus 3pt minus 5pt
\parskip=3pt plus 1.5pt
\setbox\strutbox=\hbox{\vrule height10pt depth4pt width0pt}
\let\sc=\ninerm
\let\big=\twelvebig \normalbaselines\rm}
\def\twelvebig#1{{\hbox{$\left#1\vbox to10pt{}\right.\n@space$}}}

\def\tenpoint{\def\rm{\fam0\tenrm}
\textfont0=\tenrm \scriptfont0=\sevenrm \scriptscriptfont0=\fiverm
\textfont1=\teni \scriptfont1=\seveni \scriptscriptfont1=\fivei
\textfont2=\tensy \scriptfont2=\sevensy \scriptscriptfont2=\fivesy
\textfont3=\tenex \scriptfont3=\tenex \scriptscriptfont3=\tenex
\def\it{\fam\itfam\tenit} \textfont\itfam=\tenit
\def\sl{\fam\slfam\tensl} \textfont\slfam=\tensl
\def\bf{\fam\bffam\tenbf} \textfont\bffam=\tenbf
\scriptfont\bffam=\sevenbf \scriptscriptfont\bffam=\fivebf
\def\tt{\fam\ttfam\tentt} \textfont\ttfam=\tentt
\def\ss{\fam\ssfam\tenss} \textfont\ssfam=\tenss
\tt \ttglue=.5em plus .25em minus .15em
\normalbaselineskip=12pt
\abovedisplayskip=12pt plus 3pt minus 9pt
\belowdisplayskip=12pt plus 3pt minus 9pt
\abovedisplayshortskip=0pt plus 3pt
\belowdisplayshortskip=7pt plus 3pt minus 4pt
\parskip=0.0pt plus 1.0pt
\setbox\strutbox=\hbox{\vrule height8.5pt depth3.5pt width0pt}
\let\sc=\eightrm
\let\big=\tenbig \normalbaselines\rm}
\def\tenbig#1{{\hbox{$\left#1\vbox to8.5pt{}\right.\n@space$}}}
\let\rawfootnote=\footnote \def\footnote#1#2{{\rm\parskip=0pt\rawfootnote{#1}
{#2\hfill\vrule height 0pt depth 6pt width 0pt}}}

\def\tenfoot{\tenpoint\hskip-\parindent\hskip-.1cm}

\overfullrule=0pt
\twelvepoint
\def\sbullet{\raise.2em\hbox{$\scriptscriptstyle\bullet$}}
\nofirstpagenotwelve
\hsize=16.5 truecm
\baselineskip 15pt

\def\ft#1#2{{\textstyle{{#1}\over{#2}}}}

\def\phys{\big|{\rm phys}\big\rangle}

\def\del{\partial}

\def\.{\,\,,\,\,}

\def\a{\alpha_0}

\def\ket#1{\big| #1\big\rangle}

\oneandahalfspace
\rightline{CTP TAMU--30/92}
\rightline{April 1992}

\vskip 2truecm
\centerline{\bf A Review of $W$ Strings}
\vskip 1.5truecm
\centerline{C.N. Pope\footnote{$^\star$}{\tenfoot Supported in part by the
U.S. Department of Energy, under
grant DE-FG05-91ER40633.}\footnote{}{\tenfoot Contribution to the
proceedings of the International Symposium on Blackholes, Wormholes, Membranes
\nl  \indent and Superstrings, HARC, The Woodlands, Texas, January 1992.}}

\vskip 1.5truecm

\centerline{\it Center for Theoretical Physics, Texas A\&M University,}
\centerline{\it College Station, TX 77843--4242, USA.}

\vskip 1.5truecm
\AB\singlespace
           We review some of the recent developments in the construction of
$W$-string theories.  These are generalisations of ordinary strings in which
the
two-dimensional ``worldsheet'' theory, instead of being a gauging of the
Virasoro algebra, is a gauging of a higher-spin extension of the Virasoro
algebra---a $W$ algebra.  Despite the complexity of the (non-linear) $W$
algebras, it turns out that the spectrum can be computed completely and
explicitly for more or less any $W$ string. The result is  equivalent to a set
of spectra for Virasoro strings with unusual central charge and intercepts.
\AE\oneandahalfspace

\vskip 2truecm
\centerline{Available from hep-th/9204093}

\np
\noindent
{\bf 1. Introduction}
\bigskip

     Ordinary string theory has its origins in two-dimensional gravity, which
may be thought of as the gauge theory of the Virasoro algebra.  The holomorphic
and anti-holomorphic general-coordinate symmetries of the two-dimensional
action for a set of free scalar fields $X^\mu$, generated by the
energy-momentum
tensor components $T$ and $\bar T$, may be made into local symmetries by
introducing gauge fields $h$ and $\bar h$ with Noether couplings to $T$ and
$\bar T$.  The fields $h$ and $\bar h$, together with an additional field (an
overall Weyl scaling factor) that may be introduced because of the Weyl
invariance of the theory, constitute the components of the two-dimensional
metric tensor.  Upon quantisation, ghosts must be introduced for the
integrations over the gauge fields $h$ and $\bar h$, leading to the conclusion
that quantum anomalies in the local Virasoro symmetries will be avoided
provided that the energy-momentum tensors $T$ and $\bar T$ have central
charges $c=26$.  The easiest way to achieve this is to take 26 scalars $X^\mu$,
which acquire the interpretation of being coordinates on a 26-dimensinal target
spacetime. In the BRST description, the anomaly-freedom condition $c=26$ for
the bosonic string arises from the requirement that the BRST operator $Q$ be
nilpotent.

     At the classical level, the equations of motion for the gauge fields $h$
and $\bar h$ imply that $T$ and $\bar T$ vanish.  At the quantum level,
these become operator conditions on physical states, necessary for the
vanishing of the expectation values of the operators $T$ and $\bar T$.  In
particular, the BRST invariance of states implies that the BRST operator should
annihilate the product of the physical and ghost vacua.  From the structure of
the ghost vacuum, one deduces that the zero modes $L_0$ and $\bar L_0$ of the
energy-momentum tensors $T$ and $\bar T$ should have eigenvalue 1 on
physical states.  Thus the physical-state conditions for the bosonic string are
$$
\eqalign{
L_0\phys=\phys,\qquad L_n\phys=0,\ n\ge1,\cr
\bar L_0\phys=\phys,\qquad \bar L_n\phys=0,\ n\ge1,\cr}\eqno(1.1)
$$
{}From these conditions, one derives the spectrum of states in the bosonic
string.

     The construction of $W$-string theories proceeds in a very similar
manner.  Let us, for now, focus on the purely holomorphic sector of such a
theory.  One introduces gauge fields $h\, B,\ldots$ for each of the
higher-spin currents $T,\, W,\ldots$ in the algebra.  Anomaly freedom of the
local $W$ symmetries at the quantum level is again achieved by requiring that
the BRST operator be nilpotent.  This ensures that one has an anomaly-free
theory of $W$ gravity [1,2].  Then BRST invariance of states in the theory
determines the intercepts $\omega_s$ for the zero modes of the matter currents,
leading to physical-state conditions
$$
V^{(s)}_0\phys=\omega_s\phys,\qquad V^{(s)}_n\phys=0,\ n\ge1,\eqno(1.2)
$$
where $V^{(s)}_n$ denotes the $n$'th Laurent mode of the spin-$s$ current
$V^{(s)}(z)$ in the algebra.  There are also analogous conditions for the
anti-holomorphic sector of the theory.

     In the following sections, we shall review some of the recent results
that have been obtained in the study of $W$-string theories, and explain how
the steps that have been outlined above are implemented in practice.  The
most remarkable feature, perhaps, is that despite the great complexity of
the $W$ algebras, one can in fact obtain complete results for the physical
spectrum in almost all cases.  A key reason for this is that it is never
necessary to know the explicit form of the algebra itself---it is sufficient
to know how it may be realised in terms of free fields.  Basic realisations
are known in terms of Miura transformations, and from these, more general
ones can be obtained that admit string-like interpretations.  Further
simplifications of the apparently formidable task of constructing $W$-string
theories are achieved by observing that one need not struggle to construct
{\it primary} higher-spin currents, although such exist (the currents that
come directly from the Miura transformation are not primary), since the
solutions to the physical-state conditions (1.2) are independent of the
choice of basis for the currents.  Finally, the daunting task of
constructing the BRST operator, for the purpose of determining  the
intercepts $\omega_s$ in (1.2), can be avoided by using a known physical
state to read off the intercepts from (1.2).  As a consequence of these
points, it turns out that the spectrum of $W$-strings can be computed
completely.  Thus $W$ strings represent a relatively modest and calculable
extension of ordinary string theory.  In this respect, they contrast rather
markedly with another class of possible extensions of string theory---$p$
branes---for which the prospects of exact quantum results seem slender.  In
fact there is a rather close connection between $W$ algebras and simple Lie
algebras, with the Virasoro algebra being associated with $su(2)$, and the
extended $W$ algebras being associated with larger simple groups.  The study
of all $W$ strings rather than merely the Virasoro string is therefore
analogous to undertaking a systematic study of all simple Lie groups rather
than focussing attention only on $su(2)$.

\bigskip\bigskip
\noindent{\bf 2. The Physical-state Conditions}
\bigskip

     The construction of the $W$-gravity theory that forms the basis of the
corresponding $W$-string theory is made completely straightforward by making
use of the BRST quantisation procedure.  We shall not discuss the details
here; they may be found, for the example of $W_3$ gravity, in [1,2].  The
generalisation to any $W$-gravity theory is immediate.  The upshot is that
one obtains an anomaly-free $W$-gravity provided that the BRST operator for
the corresponding $W$ algebra is nilpotent.

     Already for $W_3$, the explicit construction of the BRST operator $Q$
is quite involved [3,4], and in fact this is the only example of a $W$ algebra
for which it has been found.  Despite the non-linearities of the $W$
algebras, however, one can derive what the central charge that will ensure
nilpotence of $Q$ must be in the following way.  The potential obstructions
to nilpotence come from the central terms in the $W$ algebra; these will
give rise to a set of non-vanishing terms with different structures in the
expression for $Q^2$. However, there is just one overall central-charge
parameter in a given $W$ algebra, and so it must be the case that when this
parameter is chosen so as to make any one anomaly structure in $Q^2$ vanish,
all the others will vanish too.  The easiest way to determine the required
central charge is to look at the anomaly structure in $Q^2$ that corresponds
to the spin-2 sector of the algebra.  Since this Virasoro subsector {\it is}
linear, one simply has to add up the contributions from the ghost pairs for
each current in the $W$ algebra in order to determine the matter central
charge that will cancel all the ghost contributions.  The ghosts for a
bosonic (fermionic) current of spin $s$ contribute $\mp 2(6s^2-6s+1)$ to the
ghostly central charge, and so the matter realisation of the $W$ algebra
must have critical central charge $c_{\rm crit}$ given by
$$
c_{\rm crit}=2\sum_{\{s\}_B}(6s^2-6s+1)-2\sum_{\{s\}_F}(6s^2-6s+1)\eqno(2.1)
$$
in order to achieve nilpotence of $Q$.  Here $\{s\}_B$ and $\{s\}_F$ denote
the sets of spins of the bosonic and fermionic currents in the $W$ algebra.

     The next step is to determine the physical-state conditions for the
corresponding $W$-string theory.  These are determined by the condition that
the product of a physical state with the ghost vacuum be invariant under the
action of the BRST operator $Q$.  This leads to the conditions (1.2), where
the only subtlety that arises is the determination of the intercepts
$\omega_s$.  These are in principle obtainable once one knows the structure
of the ghost vacuum, and the explicit form of the BRST operator.  The ghost
vacuum is quite easily constructed; the general derivation for a theory with
arbitrary numbers of bosonic and fermionic currents is given in [5].  One
starts with an $SL(2,C)$-invariant ghost vacuum.  The true ghost vacuum is
then obtained by acting on this with the appropriate product of ghost
creation operators.  The ghosts $c^{(s)},\ b^{(s)}$ for a bosonic current of
spin $s$ contribute a factor $c^{(s)}\del c^{(s)}\del^2 c^{(s)}\cdots
\del^{s-2}c^{(s)}(0)$.  This may be elegantly bosonised, using the rules
$$
b^{(s)}\to e^{-i\phi^{(s)}},\qquad c^{(s)}\to e^{i\phi^{(s)}},\eqno(2.2)
$$
and rewritten as a factor $e^{i(s-1)\phi^{(s)}}(0)$. In a similar
way, a fermionic current of spin $s$ implies the inclusion of a factor
$e^{-(s-\ft12)\sigma^{(s)}}(0)$ acting on the $SL(2,C)$ ghost
vacuum, where $\sigma^{(s)}$ is the scalar field appearing in the
bosonisation of the $\beta,\ \gamma$ ghost system for a fermionic current of
spin $s$ [5].  In general, the non-linearities of the $W$ algebra make the
explicit  construction of the BRST operator prohibitively complicated.
However,
as in  the $Q^2$ calculation described above, the linearity of the Virasoro
subalgebra implies that calculations in the spin-2 sector can be performed
quite easily.  Here, one finds that the intercept $\omega_2$ for the spin-2
zero-mode $L_0$ is simply given by the negative of the ghostly spin-2 zero
mode acting on the ghost vacuum that we have just described.  Thus one
simply has to count the ghost-number contributions from each factor acting
on the $SL(2,C)$ ghost vacuum.  The result for the spin-2 intercept is then
easily found to be
$$
\omega_2=\ft12\sum_{\{s\}_B}s(s-1) -\ft12\sum_{\{s\}_F}\big(s-\ft12\big)^2.
\eqno(2.3)
$$
This may be recast into the simple form
$$
\omega_2=\ft1{24}\big(c-2N_B-N_F\big),\eqno(2.4)
$$
where $c$ is the critical central charge given by (2.1), and $N_B$ and $N_F$
are the numbers of bosonic and fermionic currents in the $W$ algebra [5].  A
procedure for determining the higher-spin intercepts, avoiding the problem of
first constructing the BRST operator, will be described later.

\bigskip\bigskip
\noindent{\bf 3. Realisations of $W\!A_n$ Algebras}
\bigskip

     Let us turn now to the question of finding appropriate realisations of
the $W$ algebras, which will enable us to give a string-like interpretation to
the $W$-gravity theories.  The {\it ur}-realisations of $W$ algebras are
provided by the Miura transformation [6].  To be specific, let us consider for
now the case of the $W_N$ algebra.  This has one current of each spin $s$ in
the
interval $2\le s\le N$.  Actually, the $W_N$ algebra is closely associated with
the Lie algebra $su(N)\cong A_{N-1}$, and it will prove to be more convenient
to consider $W_{n+1}$.  There is in fact a procedure for associating a $W$
algebra $W\!G$ with every compact semisimple Lie group $G$, and so we shall
denote the $W_{n+1}$ algebra to be considered here by $W\!A_n$.  The Miura
transformation provides a realisation of $W\!A_n$ in terms of $n$ free scalar
fields $\vec\varphi^{(n)}=(\varphi_1,\ldots,\varphi_n)$, {\it via} the
following
construction [6]:
$$
\prod_{k=1}^{n+1}\Big(\a\del +\vec h_k^{(n)}\cdot\del\vec\varphi^{(n)}\Big) =
(\a\del)^{n+1} +\sum_{s=2}^{n+1}W_s^{(n)}(\a\del)^{n+1-s},\eqno(3.1)
$$
where $W_s^{(n)}$ denotes the spin-$s$ current of the $W\!A_n$ algebra.  The
$(n+1)$ vectors $\vec h_k^{(n)}$, which have $n$ components, satisfy the
defining relations
$$
\eqalignno{
\vec h_i^{(n)}\cdot \vec h_j^{(n)}&=\delta_{ij}-{1\over n+1},&(3.2a)\cr
\sum_{k=1}^{n+1}\vec h_k^{(n)}&=0.&(3.2b)\cr}
$$
Any set of vectors that satisfy these equations will define, {\it via}
(3.1), a set of currents that close on the $W\!A_n$ algebra.  Different
choices will yield the algebra in different bases.  For $k=1,\ldots,n$, the
$\vec h_i^{(n)}$ are the weights of the $(n+1)$-dimensional representation
of $A_n$ in some basis.  It is easy to see from (3.2$a,b$) that they may be
defined recursively as
$$
\vec h_i^{(n)}=\big(\vec h_i^{(n-1)},\ft1{\sqrt{n(n+1)}}\big).\eqno(3.3)
$$
This recursive definition will prove crucial in what follows.

     Defining $\phi_n\equiv \ft1{\sqrt{n(n+1)}}\varphi_n$, and making use of
(3.3), it follows that the left-hand side of the Miura transformation may
be written as [7]
$$
\Big(\a\del -n\del\phi_n\Big)e^{-\phi_n/\a}\prod_{k=1}^n\Big(\a\del +\vec
h_k^{(n-1)}\cdot\del\vec\varphi^{(n-1)}\Big)e^{\phi_n/\a}.\eqno(3.4)
$$
But the product of operators sitting between the exponentials in (3.4) is
precisely the left-hand side of the Miura transformation (3.1) for the
$W\!A_{n-1}$ algebra, and so we see that using (3.1) the currents $W^{(n)}_s$
of $W\!A_n$ may be written in terms of the currents $W^{(n-1)}_s$ of the
$W\!A_{n-1}$ algebra together with one free scalar field $\phi_n$.  The
explicit expressions are [7]:
$$
\eqalign{
W^{(n)}_k=\sum_{q=0}^k &{n+1+q-k\choose q}\Big[ {n+1-k\over n+1+q-k}
W^{(n)}_{k-q} P_q(\phi_n)\cr
&+\a\del\Big(W^{(n)}_{k-q-1} P_q(\phi_n)\Big) -n(\del\phi_n)
W^{(n)}_{k-q-1} P_q(\phi_n)\Big],\cr}\eqno(3.5)
$$
where $P_q(\phi_n)$ is defined by
$$
P_q(\phi_n)\equiv e^{-\phi_n/\a}\Big((\a\del)^q e^{\phi_n/\a}\Big).\eqno(3.6)
$$

     By applying these results recursively, we may express the
currents of $W\!A_n$ in terms of the energy-momentum tensor of
$W\!A_1\cong W_2\cong \hbox{Virasoro}$, together with $(n-1)$ additional scalar
fields $(\varphi_2,\ldots,\varphi_n)$.  Although the Miura transformation gives
the energy-momentum tensor explicitly in terms of the single free scalar
$\varphi_1$, it is clear that since all the scalars commute with each other, we
may replace the energy-momentum tensor given in terms of $\varphi_1$ by an
arbitrary one $T^{\rm eff}$, as long as it has the same central charge.  This
was first discovered for the case pf the $W_3$ algebra in [8].  In detail, we
find from (3.5) that the energy-momentum tensor for the $W\!A_n$ algebra is
given by $$
W^{(n)}_2=-\ft12 \del\vec\varphi^{(n)}\cdot\del\vec\varphi^{(n)} +\a
\vec\rho^{(n)}\cdot\del^2\vec\varphi^{(n)},\eqno(3.7)
$$
where $\vec\rho^{(n)}$ is the Weyl vector of $A_n$, given in terms of $\vec
h^{(n)}_i$ by
$$
\vec\rho^{(n)}=\sum_{j=1}^n(n+1-j)\vec h^{(n)}_j.\eqno(3.8)
$$
The central charge for this realisation of the Virasoro algebra is
$$
c_n=n\Big(1+(n+1)(n+2)\a^2\Big).\eqno(3.9)
$$
The energy-momentum tensor for $\varphi_1$ in (3.7) is
$$
-\ft12(\del\varphi_1)^2+\ft1{\sqrt2}\a\del^2\varphi_1,\eqno(3.10)
$$
which has central charge $(1+6\a^2)$.  Thus by replacing (3.10) by $T^{\rm
eff}$, we have a realisation of $W\!A_n$ in terms of an energy-momentum tensor
$T^{\rm eff}$ with central charge
$$
c^{\rm eff}=1+6\a^2,\eqno(3.11)
$$
together with $(n-1)$ additional free scalar fields
$(\varphi_2,\ldots,\varphi_n)$.  The central charge of the realisation of
$W\!A_n$ is given by (3.9).

\bigskip\bigskip
\noindent{\bf 4. $W\!A_n$ String Theory}
\bigskip

     The strategy for obtaining a $W\!A_n$ string theory is now as follows.  We
first choose a realisation of the energy-momentum tensor $T^{\rm eff}$ in terms
of scalar fields $X^\mu$.  The critical central charge condition (2.1), applied
to the case of the string based on the $W\!A_n$ algebra, gives $c_{\rm
crit}=2n(2n^2+6n+5)$.  This implies from (3.9) that the background-charge
parameter $\a$ must take its critical value, given by
$$
(\a)^2={(2n+3)^2\over (n+1)(n+2)}.\eqno(4.1)
$$
{}From (3.11), we see then that the energy-momentum tensor $T^{\rm eff}$ must
have
central charge given by
$$
c^{\rm eff}=26-\Big(1-{6\over (n+1)(n+2)}\Big).\eqno(4.2)
$$
Since, for $n\ge 2$, this is not an integer, it follows that to realise $T^{\rm
eff}$ in terms of scalar fields $X^\mu$, we must include a background charge.
Without loss of generality, the coordinates $X^\mu$ of the target spacetime may
be oriented such that the background-charge vector is aligned along a
particular coordinate direction.

     The physical states of the theory are defined to be those that satisfy the
conditions (1.2).  The intercept $\omega_2$ for the $W\!A_n$ string can be read
off from the general results (2.3) or (2.4).  The result is
$$
\omega_2=\ft16 n(n+1)(n+2). \eqno(4.3)
$$
As mentioned earlier, a direct computation of the higher-spin intercepts by
acting with the BRST operator on the product of a physical state with the ghost
vacuum is prohibitively complicated, because of the difficulties of explicitly
constructing the BRST operator for such non-linear algebras.  Instead we may
exploit an observation first made in [9], and subsequently developed in [7].
It
was noticed in [9] that in the case of the $W\!A_2\cong W_3$ algebra, for the
2-scalar {\it ur}-realisation from the Miura transformation, the known
spin-2 and spin-3 intercepts implied that amongst the tachyonic states,
described by acting on the vacuum with operators of the form
$\exp(\vec\beta\cdot\vec\varphi)$, is a particular state for which the momentum
$\vec\beta$ is a certain multiple of the Weyl vector $\vec\rho$.  It was
conjectured that for {\it all} $W$ strings, such a physical state should exist.
The corresponding operator
$$
\exp(\lambda\a\vec\rho\cdot\vec\varphi)\eqno(4.4)
$$
is known as the ``cosmological operator.'' Since $W_3$ is the only example
for which the BRST calculation of the full set of intercepts has been
performed [3], no other direct test for the conjecture is available.
Assuming the conjecture is correct, it is then easy to determine all the
higher-spin intercepts:  The unknown constant of proportionality $\lambda$
in $\vec\beta=\lambda\a\vec\rho^{(n)}$ can be calculated from the knowledge
of just one intercept.  Since we do have an explicit expression for
$\omega_2$, this provides the necessary piece of information that determines
$\lambda$, and hence all the higher-spin intercepts can now be read off by
applying the zero-modes of the currents to the putative physical state. The
{\it lacuna} in this argument is the understanding of why the ``cosmological
operator'' should always give a physical tachyonic state.  The best evidence
in support of the conjecture comes from the study of $W_N$ strings, where it
has been shown explicitly for the cases of the $W_3$, $W_4$, and $W_5$
strings that any values for the intercepts other than those that follow from
this conjecture will give rise to a theory that is non-unitary [7].  It
seems highly plausible that this feature will persist in all $W$-string
theories.  It would be very interesting to obtain a proof of this important
conjecture.

     Proceeding for now under the assumption that the cosmological operator is
physical, it follows from (4.3) that the constant $\lambda$ in (4.4) is given
by
$$
\lambda_\pm=\Big(1\pm{1\over 2n+3}\Big).\eqno(4.5)
$$
Without loss of generality, we may choose the $+$ sign in (4.5), and define
this to be the ``cosmological solution.''  Substituting (4.4) and (4.5) into
the
physical-state conditions (1.2) now enables one in principle to calculate all
the higher-spin intercepts $\omega_s$.  The first few examples are given in
[7].

     The process of calculating the physical spectrum of the $W\!A_n$ string is
now straightforward in principle, since we have realisations in terms of scalar
fields $\vec\varphi$ and $X^\mu$, and the criticality and intercept conditions
for physical states are determined.  It turns out that the physical states of
the theory can be described in the following way.  Let us first consider the
situation when we work with the {\it ur}-realisation of $W\!A_n$ in terms of
the $n$ scalars $\vec\varphi$.  At the tachyonic level, the physical operators
will be of the form
$$
e^{\vec\beta\cdot\vec\varphi}.\eqno(4.6)
$$
The physical-state conditions for tachyons will clearly simply be the zero-mode
intercept conditions in (1.2).  The spin-$s$ condition will give a polynomial
of degree $s$ in $\vec\beta$.  Thus in all, we have $n$ equations, coming
from the intercept conditions for the currents for spins 2 up to $n+1$, on the
$n$ components of $\vec\beta$.  Since the conditions are all independent, it
follows that the solutions will comprise a discrete set of values for
$\vec\beta$.  The number of such discrete solutions is given by the product of
the degrees of the polynomial equations, and so there will be $(n+1)!$
solutions for the $W\!A_n$ string.  Amongst them, of course, is the
``cosmological solution''  (4.4) with $\lambda=\lambda_+$ given by (4.5).
Defining a shifted momentum vector $\vec\gamma$ by
$$
\vec\beta=\vec\gamma+\a\vec\rho,\eqno(4.7)
$$
it is straightforward to show that the system of polynomials in $\vec\gamma$
coming from the physical-state conditions (1.2) has a discrete symmetry [6],
under the action of the Weyl group of $A_n$ [7].  This group has dimension
$(n+1)!$, and it acts transitively on the Weyl vector $\vec\rho$.  Thus in
fact the discrete set of $(n+1)!$ solutions of the tachyonic physical-state
conditions is generated by taking the cosmological solution itself, and its
Weyl-reflected partners.  This is the complete set of physical states at the
tachyonic level.

     Now let us consider the more physically-interesting situation where we
replace the energy-momentum tensor (3.10) for $\varphi_1$ by an arbitrary one
$T^{\rm eff}$ built from additional scalars $X^\mu$.  It turns out now that the
momentum components $(\beta_2,\ldots,\beta_n)$ continue to be frozen to the
same sets of values that occurred in the $n$-scalar realisation [7].
Consequently, the tachyonic physical states will have the form
$$
\ket{p}=e^{\beta_2\varphi_2+\cdots\beta_n\varphi_n}(0) \ket{p}_{\rm
eff},\eqno(4.8)
$$
where $\ket{p}_{\rm eff}$ describes the effective tachyonic state in the
spacetime with coordinates $X^\mu$, and satisfies the effective
physical-state conditions
$$
L^{\rm eff}_0\ket{p}_{\rm eff}=\omega_2^{\rm eff}\ket{p}_{\rm eff},\qquad
L^{\rm
eff}_n \ket{p}_{\rm eff} =0,\ n\ge1,\eqno(4.9)
$$
where $L^{\rm eff}_n$ are the Laurent modes of
$$
T^{\rm eff}=-\ft12\eta_{\mu\nu}\del X^\mu \del X^\nu +\alpha_\mu\del^2 X^\mu,
\eqno(4.10)
$$
with $\mu=0,1,\ldots,D-1$, and the background-charge vector $\alpha_\mu$
chosen so that the central charge $D+12\alpha_\mu\alpha^\mu$ is equal to the
value $c^{\rm eff}$ given in (4.2), which was required for criticality of the
$W\!A_n$ algebra.  Since $T^{\rm eff}$ replaces the energy-momentum tensor
(3.10) for $\varphi_1$, it follows that the effective spin-2 intercept values
$\omega_2^{\rm eff}$ are given in terms of the frozen $\beta_1$ values found
above for the $n$-scalar {\it ur}-realisation. In turn, these may be found by
acting with the Weyl group on the shifted momentum $\vec\gamma$ corresponding
to
the ``cosmological'' solution. The result turns out to be [7]
$$
\omega_2^{\rm eff}=1-{k^2-1\over 4(n+1)(n+2)},\eqno(4.11)
$$
where $k$ is an integer lying in the range $1\le k\le n$.

     The above discussion generalises to higher-level states.  We must divide
these into two categories.  The first consists of states for which the
excitations take place exclusively in the unfrozen directions $X^\mu$.  The
second category consists of states where frozen directions are excited too.
For reasons that we shall explain later, it turns out that no states in the
latter category can be physical.  Thus we shall concentrate on the first
category of excited states for now.  These may be written as
$$
\ket{\rm phys}=e^{\beta_2\varphi_2+\cdots\beta_n\varphi_n}(0) \ket{\rm
phys}_{\rm eff},\eqno(4.12)
$$
where again the values of $\beta_2,\ldots\beta_n$ are the same set of frozen
values found in the $n$-scalar tachyon calculation.  The physical-state
conditions (1.2) imply that the effective physical states $\ket{\rm phys}_{\rm
eff}$ must satisfy the effective physical-state conditions
$$
L^{\rm eff}_0\ket{\rm phys}_{\rm eff}=\omega_2^{\rm eff}\ket{\rm phys}_{\rm
eff},\qquad L^{\rm eff}_n \ket{\rm phys}_{\rm eff} =0,\ n\ge1,\eqno(4.13)
$$
where again the effective intercept values $\omega_2^{\rm eff}$ are given by
(4.11).  The effective physical states $\ket{\rm phys}_{\rm eff}$ are built up
just as in ordinary string theory, by acting on an effective level-0 state with
appropriate operators constructed from products and derivatives of $\del X^\mu$
contracted into polarisation tensors.

     The conclusion of the above discsussion is that the $W\!A_n$ string has a
physical spectrum that is the same as that for a set of Virasoro-like string
theories, with the non-standard value $c^{\rm eff}$ for the central charge,
given by (4.2), and the set of non-standard values $\omega_2^{\rm eff}$ for the
intercepts, given by (4.11) [7].  Note that the case $k=1$ in (4.11) gives an
effective intercept $\omega_2^{\rm eff}=1$, which implies that this sector of
the theory gives the same mass spectrum as the Virasoro string.  In particular,
it describes the usual massless states of string theory.  Results for the case
of the $W\!A_2\cong W_3$ string were found in [10].

     The momentum-freezing phenomenon that we met above has a number of
important consequences, so we shall now consider this in a
bit more detail.  A full discussion of the relevant issues may be found in
[11].
Although we are calling $\vec\beta$ the ``momentum'' in the $\vec\varphi$
directions, it is clear from (4.6) that the true momentum is really
$-i\vec\beta$.  Since the physical-state conditions freeze $\beta$ to {\it
real}
sets of values, it follows that the true momentum is frozen to imaginary
values.  On the face of it, this sounds rather disturbing, since we are
acustomed to dealing only with real momenta in ordinary physics.  One way of
seeing why momentum normally must be real is that if one calulates correlation
functions such as $\int dx e^{-ip'x} e^{ipx}$, the integral will only be
well-defined ({\it qua} distribution) if the imaginary part of $p'-p$ vanishes;
otherwise, the integrand would diverge exponentially.  Since it must be well
defined for any pair of states one wishes to consider, it follows that $p$ must
be real for any state.  The situation is different in our case, since we have a
background-charge vector $\a \vec\rho$ in the energy-momentum tensor (3.7),
which can be viewed as an injection of momentum $-i\a\vec\rho$ at infinity.
Thus to make correlation functions well defined, it is now necessary that each
state have a momentum whose imaginary part is precisely tuned to subtract out
the imaginary contribution from the background charge.  In terms of
$\vec\beta$,
this means that correlation functions can be made well defined if the frozen
components $(\beta_2,\ldots,\beta_n)$ have {\it real} parts equal to
$(\a\rho_2,\ldots, \a\rho_n)$. But from (4.7), this means that the
components $(\gamma_2,\ldots,\gamma_n)$ of the shifted momentum would all have
to vanish (since, when non-zero, they are always real).  However, this can only
happen in the special case of the $W\!A_2$ string; it is a general result from
group theory that at most one component of the vector obtained by acting on the
Weyl vector of $A_n$ with the Weyl group can be zero.

     Fortunately, there is another way in which the exponential divergence of
the integrand in correlation functions can be avoided.  In a closed $W$ string,
there will be both left-moving and right-moving fields, including those
corresponding to the frozen directions.  It is possible to arrange for the
imaginary momentum from the background charge to be cancelled by tuning the
{\it
total} imaginary contribution from the left-moving and right-moving sectors for
any state appropriately [11].  This works easily and naturally for any $W\!A_n$
string, since a particular consequence of the Weyl-group symmetry is that if
$\vec\gamma$ solves the physical-state conditions, then so does $-\vec\gamma$.
Thus the left-moving and right-moving momenta for the frozen fields of any
state
can be paired together, so that the total momentum does exactly cancel the
imaginary momentum at infinity.  Thus we have the interesting situation
that although there are equal numbers of left-moving and right-moving frozen
fields $(\varphi_2,\ldots,\varphi_n)$
and $(\tilde\varphi_2,\ldots,\tilde\varphi_n)$, they have to have unequal
momenta.  Since the frozen fields are therefore being treated heterotically,
they cannot be assembled into coordinates.  Of course the ``coordinates'' would
in any case have been frozen, and unobservable, so it is somewhat academic
whether they are left-right symmetric or not.  Finally, let us return to the
consideration of higher states with excitations in the frozen directions.  For
these it turns out that the $\vec\gamma\to-\vec\gamma$ symmetry that played a
crucial r\^ole above no longer occurs.  Consequently, one cannot build such
states with well-defined correlation functions, and so they will not occur in
the physical spectrum [11,7].

\bigskip\bigskip
\noindent{\bf 5. Generalisations and Discussion}
\bigskip

     We have seen in the previous section that the physical-state conditions
for the $W\!A_n$ string can be reduced to a set of effective physical-state
conditions for Virasoro-like strings.  These effective string theories have a
non-standard central charge $c^{\rm eff}$, given by (4.2), and a set of
non-standard spin-2 intercept values, given by (4.11).  A striking feature of
the expression (4.2) for the effective central charge is that it is equal to
26, the critical central charge for the bosonic string, minus the central
charge
of the $(n+1,n+2)$ Virasoro minimal model.  Furthermore, the effective
intercept
values (4.11) can be written as $\omega_2^{\rm eff}=1-\Delta_{k,k}$, where 1 is
the intercept for the usual Virasoro string and $\Delta_{r,s}$ are the
dimensions of the primary fields of the $(n+1,n+2)$ minimal model. These
observations were first made for the $W\!A_2$ and $W\!A_3$ strings in [9], and
devloped for $W\!_n$ strings in [7]. The underlying significance of this
connection between $W\!A_n$ strings and minimal models remains somewhat
obscure.

     The unitarity of the physical spectrum of the $W\!A_n$ string can be
studied by investigating the unitarity of the spectra of the effective Virasoro
string theories.  Since these all have an effective central charge which is
less
than 26, it follows that there will be a range of effective spin-2 intercept
values for which unitarity is achieved.  In fact this range is precisely
spanned by the discrete set of effective intercepts given by (4.11), and so we
see that the $W\!A_n$ string theories are all unitary [7].

     The construction of $W\!A_n$ strings that we have described here can be
extended to other $W$ algebras as well.  The essential ingredient is that one
should be able to find realisations of the algebras in terms of an
energy-momentum tensor together with additional fields.  This can be done for
the $W\!D_n$ and $W\!B_n$ algebras, based on the Lie algebras $D_n$ and
$B_n$.  As for the case of $W\!A_n$, the key point is that the Miura
transformations [12,13] can be factorised into a product of transformations
for a subalgebra, leading to the above realisations by iteration [14].  In the
case of $W\!B_n$, there are realisations in terms of an $N=1$ super-Virasoro
energy-momentum tensor together with additional fields, and so one obtains a
description in terms of effective $N=1$ superstring theories [14]. Connections
with minimal models arise in all cases.

     Further generalisations are also possible in which one realises the $W$
algebra in terms of more than one arbitrary energy-momentum tensor.  It was
first realised that realisations with two energy-momentum tensors are possible
[15], and subsequently this was generalised to multiple energy-momentum tensors
[16,17].  The key point here is that for any of the $A_n$, $D_n$ or $B_n$
series
one can reduce to a product subalgebra by deleting any vertex in the Dynkin
diagram, and factorise the Miura transformation for the original algebra into
the product of Miura transformations for the two factors in the subalgebra
[16,17].  By applying this iteratively, realisations in terms of many arbitrary
energy-momentum tensors can be found [17].  These would give rise to string
theories with multiple ``spacetimes.''

     We have seen that the central charge $c^{\rm eff}$ for the effective
energy-momentum tensor $T^{\rm eff}$ is always non-integer for non-trivial $W$
algebras; for example it is given by (4.2) for the $W\!A_n$ string, with $25<
c^{\rm eff}\le 25\ft12$.  Because of this, the background-charge vector
$\alpha_\mu$ in (4.10) must always be non-zero.  This leads to a breaking of
the
$D$-dimensional Poincar\'e group.  Assuming that we take $\alpha_\mu$ to lie in
a spacelike dimension, then it must be {\it imaginary} if the number $D$ of
spacetime dimensions exceeds 25.  As discussed in [10,7], this implies that
the coordinate direction parallel to the background-charge vector appears as a
{\it periodic} variable in the functional integral, and hence it is
automatically compactified on a circle.  Since the radius is of Planck size,
one is left, {\it \`a la} Ka\l uza Klein, with a $(D-1)$-dimensional
observable spacetime with Poincar\'e invariance in the directions orthogonal to
the background-charge vector.

    The study of $W$ string theories open up many new lines of investigation.
The intriguing relation with minimal models deserves further attention.  So
far, most of the effort has been concentrated on obtaining realisations of the
algebras, and then determining the spectrum of physical states.  One of the
outstanding problems at present is to try to build in interactions into the
theories.  Work on this problem is in progress.

\bigskip\bigskip
\centerline{\bf Acknowledgments}
\bigskip

     I am very grateful to my collaborators in the work described in this
review, namely Hong Lu, Larry Romans, Stany Schrans, Ergin Sezgin, Kelly
Stelle,
Xujing Wang and Kaiwen Xu.

\bigskip\bigskip

\centerline{\bf REFERENCES}
\frenchspacing
\bigskip

\item{[1]}C.N.\ Pope, L.J.\ Romans and K.S.\ Stelle, {\sl Phys.\
Lett.}\ {\bf 268B} (1991) 167.

\item{[2]}C.N.\ Pope, L.J.\ Romans and K.S.\ Stelle, {\sl Phys.\ Lett.}\ {\bf
269B} (1991) 287.

\item{[3]}J.\ Thierry-Mieg, {\sl Phys.\ Lett.}\  {\bf 197B} (1987) 368.

\item{[4]}K.\ Schoutens, A.\ Sevrin and P.\  van Nieuwenhuizen, {\sl Comm.\
Math.\ Phys.}\ {\bf 124} (1989) 87.

\item{[5]}H.\ Lu, C.N.\ Pope, X.J.\ Wang and K.W.\ Xu, ``Anomaly Freedom and
Realisations for Super-$W_3$ Strings,''  preprint CTP TAMU-85/91, to appear in
{\sl Nucl.\ Phys.\ B}.

\item{[6]}V.A.\ Fateev and S.\ Lukyanov,  {\sl Int.\ J.\ Mod.\  Phys.}\ {\bf
A3} (1988) 507.

\item{[7]}H.\ Lu, C.N.\ Pope, S.\ Schrans and K.W.\ Xu, ``The Complete Spectrum
of the $W_N$ String,''  preprint CTP TAMU-5/92, KUL-TF-92/1.

\item{[8]}L.J.\  Romans, {\sl Nucl.\  Phys.}\ {\bf B352} (1991) 829.

\item{[9]}S.R.\ Das, A.\ Dhar and S.K.\ Rama, {\sl Mod.\ Phys.\ Lett.}\
{\bf A6} (1991) 3055;\nl
``Physical states and scaling properties of $W$  gravities and $W$ strings,''
TIFR/TH/91-20.

\item{[10]}C.N.\ Pope, L.J.\ Romans, E.\ Sezgin and K.S.\ Stelle, {\sl Phys.\
Lett.\ }  {\bf 274B} (1992) 298.

\item{[11]}H.\ Lu, C.N.\ Pope and K.S.\ Stelle, ``Massless States in $W$
Strings,'' in preparation.

\item{[12]}S.L.\ Lukyanov and V.A.\ Fateev, {\sl Sov.\ Scient.\ Rev.}\ {\bf
A15} (1990);\nl {\sl Sov.\ J.\ Nucl.\ Phys.}\ {\bf 49} (1989) 925.

\item{[13]}A.\ Bilal and J.-L.\ Gervais, {\sl Nucl.\ Phys.}\ {\bf B314}
(1989) 646; {\bf B318} (1989) 579.

\item{[14]}H.\ Lu, C.N.\ Pope, S.\ Schrans and X.J.\ Wang, ``On Sibling
and Exceptional $W$ Strings,''  preprint CTP TAMU-10/92, to appear in {\sl
Nucl.\ Phys.\ B}.

\item{[15]}H.\ Lu, C.N.\ Pope, S.\ Schrans and X.J.\ Wang, ``New
Realisations of $W$ Algebras and $W$ Strings,''  preprint, CTP TAMU-15/92, to
appear in {\sl Mod.\ Phys.\ Lett.\ A}.

\item{[16]}G.M.T.\ Watts, ``A Note on $W$-algebra Realisations,''  preprint
DUR-CPT 92-15.

\item{[17]}H.\ Lu and C.N.\ Pope, ``On Realisations of $W$ Algebras,'' preprint
CTP TAMU-22/92.

\end